\documentclass{PoS}
\usepackage{url}
\usepackage{algorithm}
\usepackage{algorithmic}

\rightline{\sffamily HUPD-1004, KANAZAWA-10-07, KEK-CP-237}\vspace*{-10mm}

\title{Improving many flavor QCD simulations using multiple GPUs}

\ShortTitle{Improving many flavor QCD simulations using multiple GPUs}

\author{M. Hayakawa$^a$, 
\speaker{K.-I.~Ishikawa}$^{b}$\thanks{{\it E-mail}: \tt{ishikawa@theo.phys.sci.hiroshima-u.ac.jp}},
 Y. Osaki$^b$, S. Takeda$^c$, S. Uno$^a$, N. Yamada$^{de}$\\
$^a$ Department of Physics, Nagoya University, Nagoya 464-8602, Japan\\
$^b$ Department of Physics, Hiroshima University, Higashi-Hiroshima 739-8526, Japan\\
$^c$ School of Mathematics and Physics, College of Science and Engineering, Kanazawa University, Kakuma-machi, Kanazawa, Ishikawa 920-1192, Japan\\
$^d$ KEK Theory Center, Institute of Particle and Nuclear Studies, High Energy Accelerator Research Organization (KEK), Tsukuba 305-0801, Japan\\
$^e$ School of High Energy Accelerator Science, The Graduate University of Advanced Studies (Sokendai), Tsukuba 305-0801, Japan}

\abstract{
We accelerate many-flavor lattice QCD simulations using multiple GPUs.
Multiple pseudo-fermion fields are introduced additively and independently
for each flavor in the many-flavor HMC algorithm.
Using the independence of each pseudo-fermion field 
and the blocking technique for the quark solver, we can assign 
the solver task to each GPU card.
In this report we present the blocking technique for
the many-flavor dynamical QCD simulations.
We investigate the effect of the blocking and the acceleration with the multiple GPUs
for the Schr\"{o}dinger functional simulations 
with Wilson $SU(3)$ plaquette gauge action and $N_f=10$ Wilson fermions.
Five pseudo-fermion fields are introduced and the quark solver task 
is distributed in the ratio of 2:3 to two GPUs.
We expect a 40\% timing reduction from the single GPU case and 
have observed a 34\% timing reduction in the test simulations.
}

\FullConference{The XXVIII International Symposium on Lattice Field Theory, Lattice2010\\
		June 14-19, 2010\\
		Villasimius, Italy}

\begin{document}

\section{Introduction}

The Large Hadron Collider (LHC) experiment has begun to trap the tail of 
Higgs boson and to find the evidence of a theory beyond the standard model (SM).
Motivated from the unnatural feature of elementary scalar Higgs field, 
many models beyond the SM have been proposed and studied. 
The technicolor (TC) model is one of them and describes the origin 
of electroweak symmetry breaking without introducing elementary scalar particles. 
The TC is a scaled-up version of QCD, but should have different features 
from the simple scaled-up QCD. The most promising TC models should have 
a slowly running (=walking) coupling and a large mass anomalous dimension.
The non-perturbative feature of such models has been investigated using lattice 
technique over the last years~\cite{TCLATTCE}.
Especially the gauge theories with many fermions, which realize the walking feature,
are very attractive.
Simulating the lattice gauge theory with many dynamical fermions is, however, a heavier task
than that for QCD, since the computational cost is roughly proportional to the number 
of dynamical fermions. 
Improving the simulation algorithm with many dynamical flavors becomes more important.

In this paper we present some techniques to improve the HMC algorithm 
with many dynamical fermions using multiple GPUs. 
The application of GPU computing to lattice field theory has been introduced 
by~\cite{QCDGame} and has been studied extensively 
in recent years~\cite{GPUapp2010,GPUappReview,GPUappOLDS}.
However the parallel GPU computations with field domain decomposition,
which is usually employed for large-scale QCD simulations,
are still very challenging because there is no efficient 
device for direct communication among GPUs.
Here we restrict our attention to a single node computation 
with multiple GPU cards to accelerate the many flavor dynamical QCD simulations 
on a rather small lattice.
We develop a blocked algorithm for the HMC algorithm with many dynamical fermions 
and test the algorithm for the Schr\"{o}dinger functional (SF) simulations with
the $SU(3)$ gauge theory with ten dynamical fermions (ten-flavor QCD).
The performance is compared among the single GPU case, the dual GPU case, and the case without GPU.
The next section describes the fermionic part of the HMC algorithm in a general form. 
We find a parallelism in the force computation of the molecular dynamics of the HMC algorithm.
We present the CPU and the GPU implementations in section~\ref{sec:Imple}.
The results are shown in section~\ref{sec:Results}. 
The summary is given in the last section~\ref{sec:Summary}.

\section{Many-flavor simulations}
The fermion determinant in the HMC partition function is written by
\begin{equation}
    \det[D]^{N_f},
\end{equation}
where $D$ is a lattice Dirac operator and $N_f$ is the number of dynamical fermions.
For simplicity we assume that $N_f$ to be an even-number, $\det[D]=\det[D^{\dag}]$ holds, 
and the mass degenerates. 
The determinant is evaluated by introducing pseudo-fermion fields in the HMC algorithm.
There are several ways to introduce the pseudo-fermion fields, additively or multiplicatively.
Here we introduce $N_{f}/2$ pseudo-fermion fields $\phi_i$ additively.
\begin{equation}
    \det[D]^{N_f} = 
\int\prod_{i=1}^{N_f/2}
{\cal D}\phi_i^{\dag}{\cal D}\phi_i
e^{-\sum_{i=1}^{N_f/2}|D^{-1}\phi_i|^{2}}.
\label{eq:EffFermion}
\end{equation}
In this form, $N_{f}/2$ $\phi_i$ fields become independent each others. 
The exponent, together with the gauge action and gauge kinetic term, 
constructs the effective action of the HMC algorithm as usual.
We need to compute the molecular dynamics (MD) force contribution 
from this effective action~(\ref{eq:EffFermion}).
The general form of the MD force is written as
\begin{equation}
    F_{\mu}(n) = \sum_{i=1}^{N_f/2} F_{\mu,i}(n),
\quad
F_{\mu,i}(n) = f[D^{-1}\phi_{i}, {D^{\dag}}^{-1}D^{-1}\phi_{i}],
\end{equation}
where $f[x,y]$ is a function of $x$ and $y$ derived from the derivative 
of the fermion action $|D^{-1}\phi_i|^{2}$ with respect to the gauge field.
By the additive introduction of the pseudo-fermion fields, 
the force contribution $F_{\mu,i}$ can be computed independently. 
We make use of this coarse-grained parallelism to employ multiple GPUs.
However it is difficult to assign the task computing fully $i$-th MD force $F_{\mu,i}$
to a single GPU because the MD force computation contains several steps
and the GPU could handle rather simple task to achieve its high efficiency.
We extract the solver part from the force computation and 
parallelize the solver part with multiple GPUs using a blocking technique.
In the next section we describe the details of extraction of the solver part
and the blocking technique.

\begin{figure}[t]
\vspace*{-2em}
\begin{minipage}[t]{0.49\hsize}
\begin{algorithm}[H]
 \caption{MD force computation in sequential version.}
 \label{alg:MDforceV1}
\begin{algorithmic}[1]
  \STATE Pseudo-fermion fields $\{\phi_i\}$ are given.
  \FOR{$i=1,\cdots,N_f/2$} 
    \STATE Solve $D x = \phi_i, \rightarrow x = D^{-1}\phi_i$.
    \STATE Solve $D^{\dag} y = x, \rightarrow y = {D^{\dag}}^{-1}x$.
    \STATE Accumulate $F_{\mu} = F_{\mu}+ f[x,y]$.
  \ENDFOR
\end{algorithmic}
\end{algorithm} 
\end{minipage}
\hfill
\begin{minipage}[t]{0.49\hsize}
\begin{algorithm}[H]
 \caption{MD force computation in blocked version.}
 \label{alg:MDforceV2}
\begin{algorithmic}[1]
  \STATE Fields $\Phi$ $=$ $(\phi_1,\phi_2,\cdots,\phi_{N_f/2})$ are given.
  \STATE Solve $D X = \Phi, \rightarrow X = D^{-1}\Phi$.
  \STATE Solve $D^{\dag} Y = X, \rightarrow Y = {D^{\dag}}^{-1}X$.
  \FOR{$i=1,\cdots,N_f/2$} 
    \STATE Accumulate $F_{\mu} = F_{\mu}+ f[x_i,y_i]$.
  \ENDFOR
\end{algorithmic}
\end{algorithm}
\end{minipage}
\end{figure}

\section{Implementation}
\label{sec:Imple}
\subsection{CPU implementation}
\label{sec:CPU}
In this subsection we describe the blocking technique and the blocked solver algorithm
employed for the CPU side computation in detail.
Alg.~\ref{alg:MDforceV1} shows the compute step of the MD force in 
the original (non-blocked) version, where the force is sequentially computed and accumulated.
This form is not suitable for the parallel execution of the solver. 
We reorganize the flavor do-loop in the blocked form
as shown in Alg.~\ref{alg:MDforceV2}. where
the working block vectors, $X = (x_1,x_2,\cdots,x_{N_f/2})$, $Y = (y_1,y_2,\cdots,y_{N_f/2})$ 
are introduced.
The linear equations are organized in the blocked form 
at the 2nd (3rd) line of Alg.~\ref{alg:MDforceV2}.

To solve the linear equations the iterative solvers, such as CG, BiCGStab, GMRES {\it etc.},
are usually employed.
The blocked form of the linear equations could have a benefit from sharing the 
Krylov subspace among $Dx_i=b_i$ and various blocked iterative solvers 
have been proposed and explored.
We implemented the Global BiCGStab (Gl-BiCGStab)~\cite{GLBICGSTAB} and 
the Blocked BiCGStab (Bl-BiCGStab)~\cite{SakuraiTadanoKuramashi}
for the CPU side double-precision solver.

Our test problem is $N_f=10$ QCD with Wilson fermions. 
We employ the site even/odd preconditioning. 
The blocked HMC algorithms are compared to the original version 
which uses the BiCGStab solver sequentially. 
On top of the blocking modification we can accelerate the blocked 
solver using multiple GPUs.

\subsection{GPU implementation}
\label{sec:GPU}
We employ GPU cards produced by Nvidia. The details of the GPU architecture 
is described in~\cite{NVIDIA}.  We write the GPU codes in the CUDA language.
To achieve high efficiency for the GPU computation the task assigned to the GPU
should be as simple as possible and should contain high parallelism.
The GPU computation has been applied to the quark solver and high efficiency has been
achieved in the case of a single GPU computation using
the mixed-precision (or flexible) preconditioner technique~\cite{GPUappReview}.
Based on these success we would like to assign one linear equation 
$Dx=b$ to one GPU even if we have many GPUs.

To accelerate the blocked solver using multiple GPUs we modify the Gl-BiCGStab to 
have the mixed-precision preconditioning functionality according to the description
in the appendix of~\cite{PACSCS}.
We control the multiple GPUs using OpenMP threading on top of the CUDA environment.
We have implicitly assumed that the number of pseudo-fermion fields equals to 
the number of GPUs until now.
If this is not the case, some GPUs may solve several linear equations sequentially or 
may solve nothing, which results in a load imbalance among GPUs.
By this modification the single-precision BiCGStab solvers on the GPUs are called
in embarrassingly parallel in the preconditioning part of the Gl-BiCGStab.
We follow the tuning techniques described in~\cite{GPUappReview} for the GPU solver.

Our test problem is $N_f=10$ QCD and five pseudo-fermion fields are introduced.
The task imbalance occurs since we have only two GPU cards. 
We distribute the task in the ratio of $2:3$ to the two GPU cards,
the one GPU solves the two equations sequentially and the other solves the three sequentially.
We also test the blocked algorithm with a single GPU
where the five equations are solved sequentially on the single GPU card.
The speed up (reduction of timing) from the single GPU case to the dual GPU case 
is expected to be $3/5=0.6$.
The single-precision coefficient matrix $D$ on the GPUs is also even/odd 
preconditioned as in the CPU side.

\subsection{CPU SSE acceleration}
\label{sec:SSE}
The GPU implementation is highly optimized and the acceleration with GPU from 
the case without GPU is almost obvious from the previous studies. 
It is fair for comparison to introduce a more tuned solver for the CPU case.
To get the best performance for the CPU case
we also implemented the single-precision solver using the SSE intrinsics 
in the C++ language and employed a more aggressive preconditioner for the Wilson fermions.
To use the multi-cores of a CPU efficiently we employed 
the locally-lexicographical site ordered SSOR (ll-SSOR) preconditioner~\cite{LLSSOR}.
To reduce the memory bandwidth requirement the 3rd column of the $SU(3)$ matrices is dropped
from the memory and is reconstructed on the fly.
The SSE intrinsics are used entirely in the single-precision solver.
Instead of the single-precision GPU solver,
the single-precision solver with the SSE and the ll-SSOR preconditioner is 
sequentially called in the preconditioning part of
the double-precision Gl-BiCGStab for each pseudo-fermion.

\section{Results}
\label{sec:Results}
We test our algorithm on a PC box which has a single CPU and two GPU cards.
The CPU is Intel's Core i7 920 (4 cores) running at 2.67 GHz, and the GPUs
are two Nvidia's GeForce GTX 285 (240 cores) cards. The OS is CentOS 5.2 (Linux).
Intel Fortran is used for the HMC algorithm and C++ is partly used for 
the SSE acceleration. The GPU code is written with the CUDA 2.3.
The whole do-loops for lattice site are parallelized with OpenMP.

We test the algorithm described above on the SF setup 
with the $SU(3)$ plaquette gauge and $N_f=10$ Wilson quarks action on a $16^4$ lattice. 
The action parameters are $\beta=4.52$ and $\kappa=0.15805$ which gives
a rather strong coupling.
To compare the solver residual history and the timing on the same basis,
we rerun the HMC algorithms (with/without various improvements)
starting from the same thermalized configuration.

\begin{figure}[h]
\begin{minipage}[t]{0.48\hsize}
    \centering
    \includegraphics[scale=0.57]{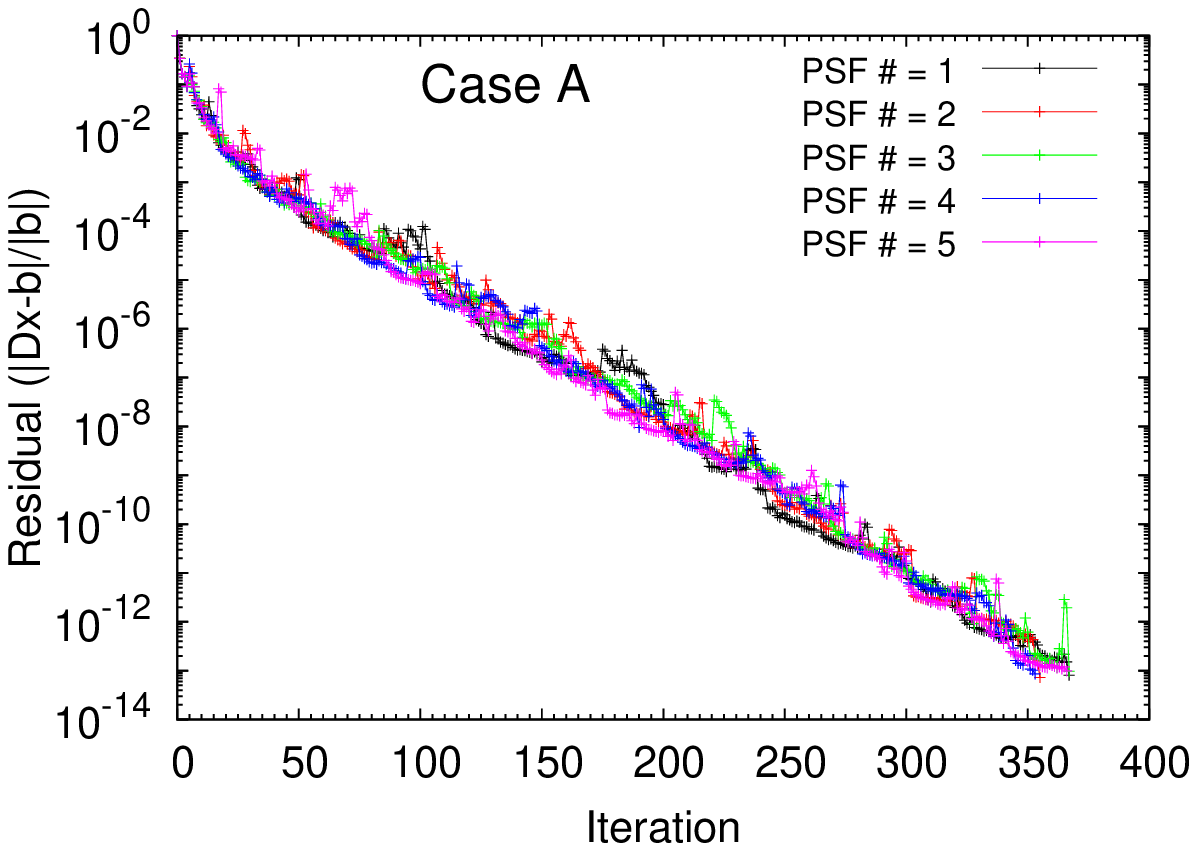}
    \vspace*{-2em}
    \caption{A sample of the residual history in the case A. 
             The five histories are overlaid for the five linear equations.}
    \label{fig:ORG}
\end{minipage}\hfill
\begin{minipage}[t]{0.48\hsize}
    \centering
    \includegraphics[scale=0.57]{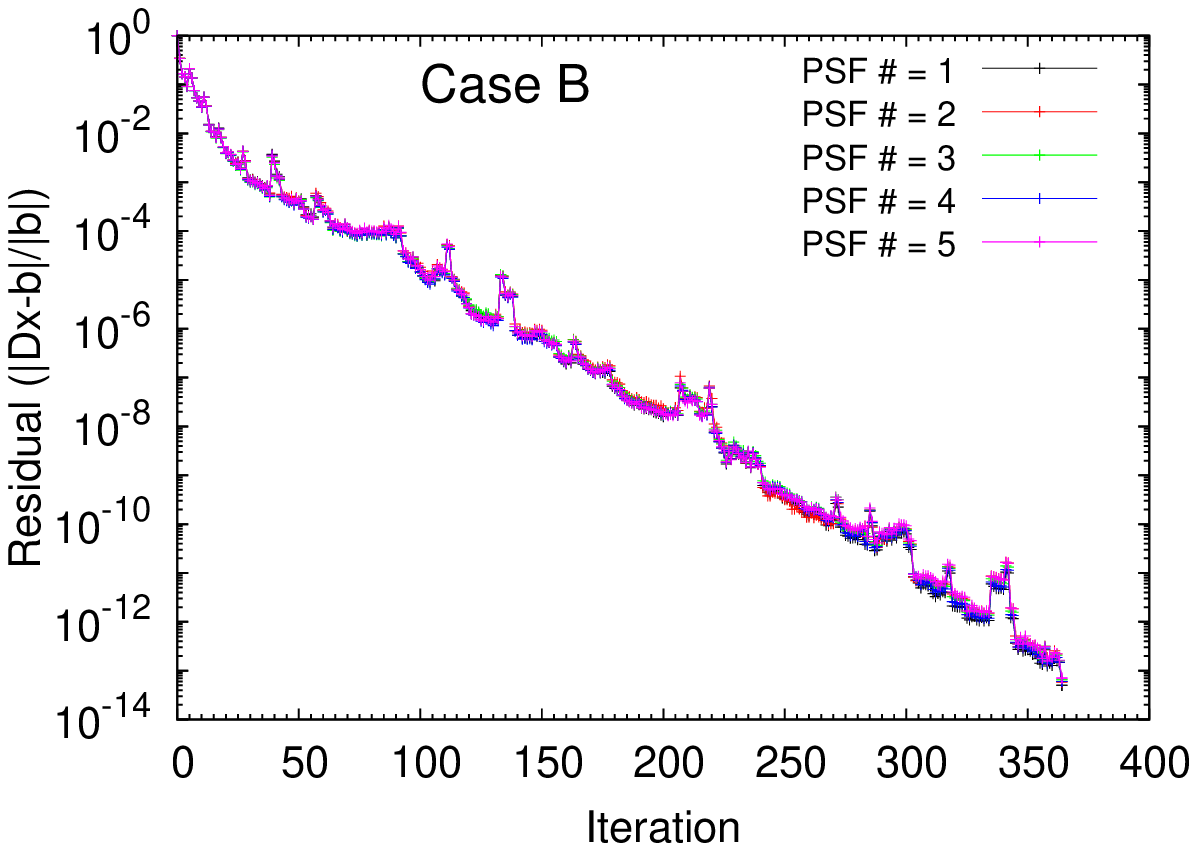}
    \vspace*{-2em}
    \caption{Same as Fig.~\protect \ref{fig:ORG}, but for the case B.}
    \label{fig:CaseB}
\end{minipage}
\end{figure}

\begin{figure}[h]
\begin{minipage}[t]{0.48\hsize}
    \centering
    \includegraphics[scale=0.57]{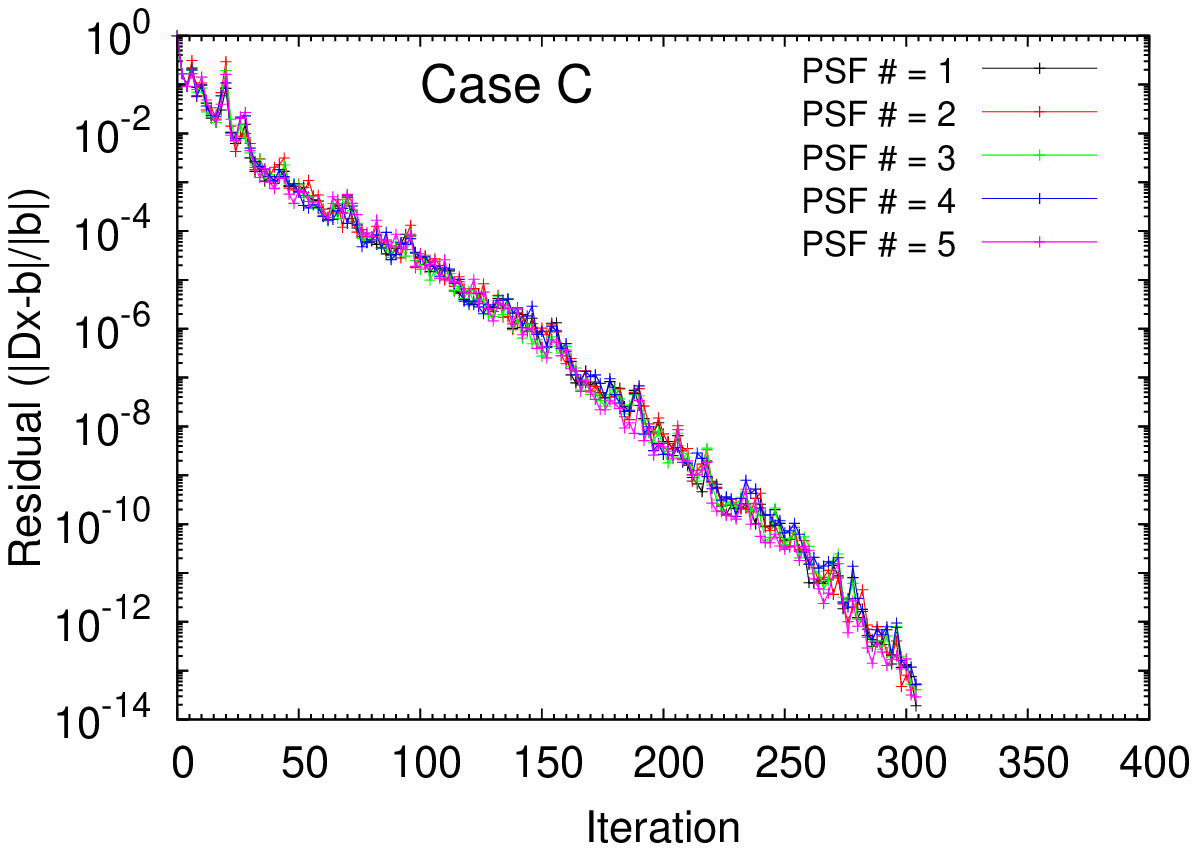}
    \vspace*{-2em}
    \caption{Same as Fig.~\protect \ref{fig:ORG} but for the case C.}
    \label{fig:CaseC}
\end{minipage}\hfill
\begin{minipage}[t]{0.48\hsize}
    \centering
    \includegraphics[scale=0.57]{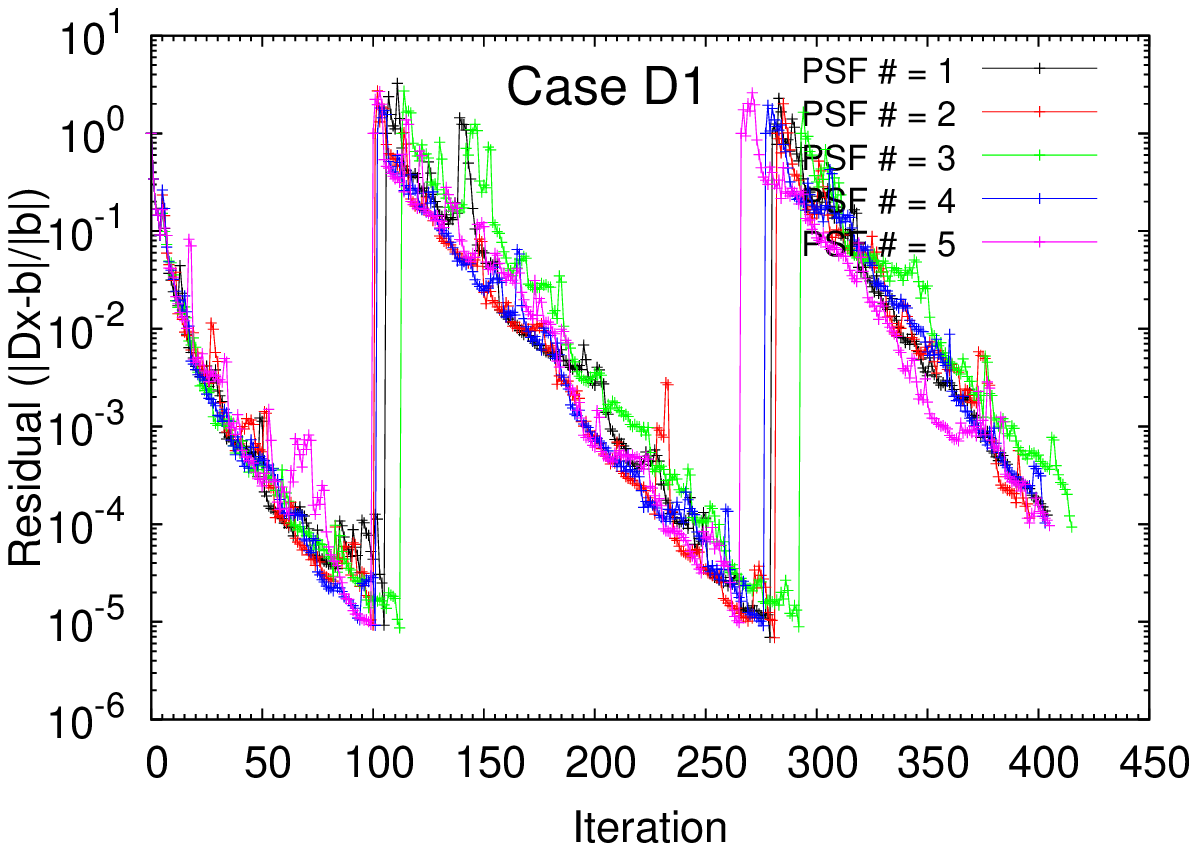}
    \vspace*{-2em}
    \caption{Same as Fig.~\protect \ref{fig:ORG}, but for the case D1. 
             The case D2 also has the same history.}
    \label{fig:CaseD1}
\end{minipage}
\end{figure}

\begin{figure}[h]
\begin{minipage}[t]{0.48\hsize}
    \centering
    \vspace*{-1em}
    \includegraphics[scale=0.57]{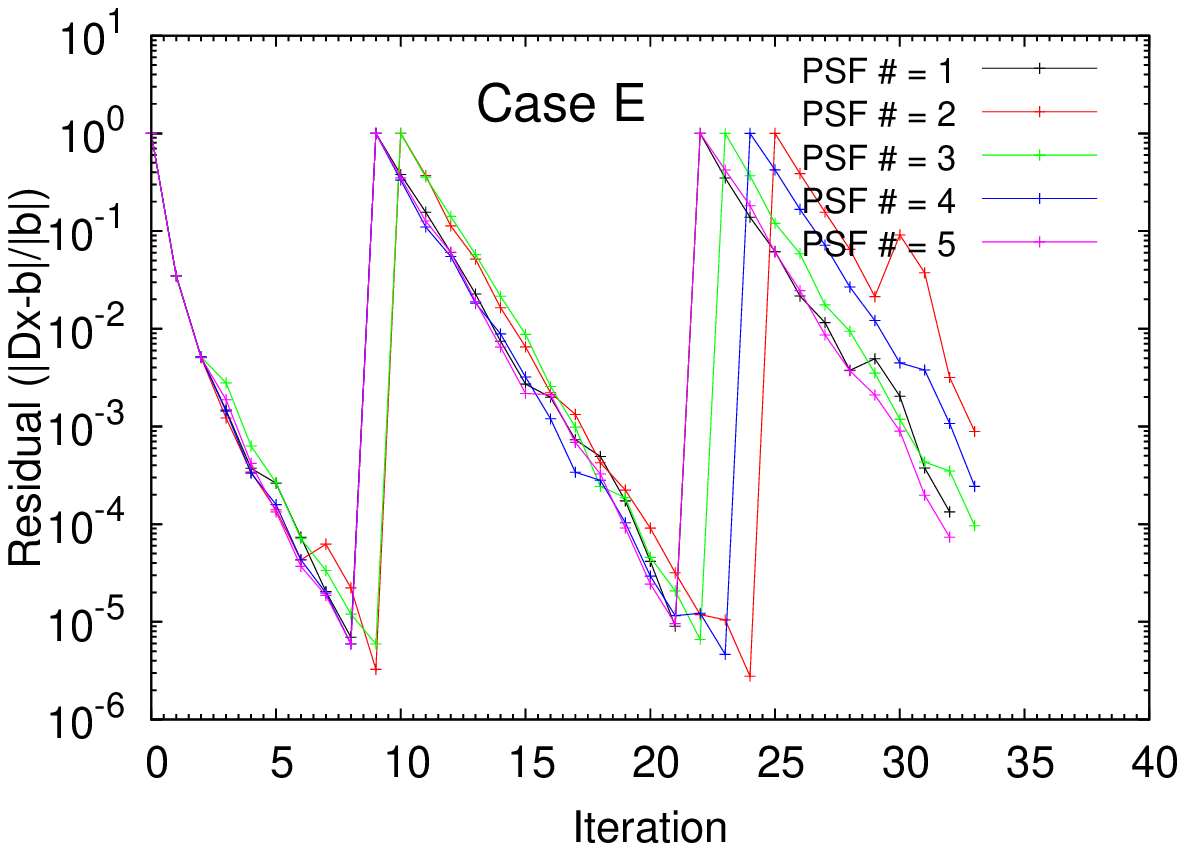}
    \vspace*{-2em}
    \caption{Same as Fig.~\protect \ref{fig:ORG}, but for the case E.}
    \label{fig:CaseE}
\end{minipage}\hfill
\begin{minipage}[t]{0.48\hsize}
   \makeatletter
   \def\@captype{table}
   \makeatother
    \centering
    \vspace*{-1em}
    \begin{tabular}{c|cc}\hline
Case  & 
\parbox[c][4.0em]{7em}{Averaged solver timing in a trajectory [sec]}  &
\parbox[c][4.0em]{6em}{HMC timing for a trajectory [sec]} \\ \hline
A  & 30.9(1.0)   & 4097 \\
B  & 25.1(0.2)   & 3346 \\  
C  & 23.1(0.1)   & 3094 \\
D1 &  4.10(0.02) &  565 \\
D2 &  2.56(0.01) &  374 \\
E  &  9.53(0.05) & 1280 \\ \hline
    \end{tabular}
    \caption{Timing comparison. 
             The averaged solver timing is for five linear equations.}
    \label{tab:timing}
\end{minipage}
\end{figure}

Table~\ref{tab:testlist} shows the test list of the combination 
of the solver algorithm, the number of GPUs, and the SSE.
The timing results are summarized in Table~\ref{tab:timing}.
A slight improvement in the solver timing is observed in the case B owing to 
the localisation of the data access and the better use of data cache by the blocking. 
The residual history for the case B (Gl-BiCGStab) is very similar to that for the case A 
and no improvement on the iteration count is observed.
The case C (Fig.~\ref{fig:CaseC}) uses the Bl-BiCGStab which shares the Krylov subspace effectively 
and we observed a 18\% reduction for the total iteration count from the case A (Fig.~\ref{fig:ORG}).
This behavior agrees with~\cite{SakuraiTadanoKuramashi}.
The timing is also reduced by 25\%.

Figure~\ref{fig:CaseD1} shows the result from the case D1. 
The case D2 also has the same history.
The history shows the residual for the single-precision GPU solver called
within the Gl-BiCGStab solver.
The GPU solver is called three times in this case to achieve 
the double-precision solutions. 
The timing is reduced by 87\% for the case D1 (single GPU case) and 
by 92\% for the case D2 (dual GPU case) from the original timing 
as shown in Tab.~\ref{tab:timing}.  
A factor of ten speed up has been observed in the literature and 
we also obtain the similar result on the timing for the single GPU case.
The speed up from the single GPU case to the dual GPU case is 62\%,
which is close to the ideal speed up of 3/5 = 60\%. 
The embarrassingly parallel execution of the GPU solver 
as a preconditioner works well.

As described in the previous section, we also implemented our best code for the CPU solver 
to make a fair comparison among the CPU and GPU computations.
Figure~\ref{fig:CaseE} shows
the residual history of the SSE single-precision-ll-SSOR preconditioned CPU solver.
The iteration count does not match with the previous figures in terms of 
the floating point number operation since four-iterations are already included in 
the SSOR preconditioner. 
The timing is also given in Tab.~\ref{tab:timing} and we observed a 69\% timing reduction. 
A naive implementation of double-precision solvers is not recommended for the Intel architecture.
One must try various improvement techniques from the algorithmic and
architectural point of view.

In the 2nd column of Tab.~\ref{tab:timing}
we tabulate the timing for a single trajectory of the HMC algorithm. The timing
is reduced by 34\% from the Case D1 to the Case D2, 
and the number deviates a little more from the ideal number 40\% according
to Amdahl's law.

\begin{table}[h]
\vspace*{-1em}
    \centering
    \begin{tabular}{c|cccc}\hline
Case  &                       & Solver               &  \# of GPUs    & SSE+ll-SSOR   \\ \hline
  A   & Non-Blocked(Original) &    BiCGStab          &   0     &   None        \\
  B   & Blocked               & Gl-BiCGStab          &   0     &   None        \\
  C   & Blocked               & Bl-BiCGStab          &   0     &   None        \\
  D1  & Blocked               & Gl-BiCGStab/BiCGStab &   1     &   None        \\
  D2  & Blocked               & Gl-BiCGStab/BiCGStab &   2     &   None        \\
  E   & Blocked               & Gl-BiCGStab/BiCGStab &   0     &    Yes        \\\hline
    \end{tabular}
    \caption{List for test case. A, B, C and E use only CPU. D1 and D2 use GPU(s).
             D1, D2, and E use the single-precision BiCGStab solver 
             with the mixed-precision technique.}
    \label{tab:testlist}
\end{table}

\section{Summary}
\label{sec:Summary}
In this paper, we have shown the blocking technique for the many-flavor dynamical QCD simulations.
The blocking can suitably distribute the solver task independently 
to many GPU cards attached to a single PC box.
We have implemented two types of the blocked quark solvers and have applied the mixed-precision 
technique with the single-precision facility of GPU and CPU. 
We have tested the blocked HMC algorithm for the Schr\"{o}dinger functional simulations
with the plaquette gauge and ten-flavor Wilson quarks.
We found the almost ideal speed up for the solver part using dual GPUs.
This algorithm has been being used to estimate the walking behavior of the coupling 
in the SF scheme partly in the strong coupling region and to search
a near conformal theory for the TC model~\cite{SFWorks}.

A part of the program development and the numerical simulations have been done on
the INSAM (Institute for Numerical Simulations and Applied Mathematics) GPU 
cluster at Hiroshima University. This work was supported in part by the Grant-in-Aid for
Scientific Research of 
Japan Society for the Promotion of Science (JSPS) and of
the Japan Ministry of Education, Culture, Sports, Science and
Technology (MEXT) (Nos. 227180, 18104005, 20105001, 20105002, 20105005, 20540261, 20740139,
21684013, 22011012, 22244018, and 22740183).

\end{document}